\documentclass[aps,twocolumn,showpacs,preprintnumbers,floats]{revtex4}
\usepackage{amsfonts}
\usepackage{amsmath}
\usepackage{graphicx}
\usepackage{dcolumn}
\usepackage{bm}

\begin{document}

\preprint{}
\title{Magnetic and dielectric properties of YbMnO$_3$ perovskite thin films}
\author{D. Rubi$^{1}$}
\email{d.rubi@rug.nl}
\author{Sriram Venkatesan$^{2}$, B.J. Kooi$^{2}$, J.Th.M. De Hosson$^{2}$,  T.T.M. Palstra$^{1}$, B. Noheda$^{1}$}
\email{b.noheda@rug.nl}
\affiliation{$^{1}$Dep. Chemical Physics, Zernike Institute for
Advanced Materials, University of Groningen, The Netherlands} \affiliation{$^{2}$Dep. Applied Physics, Zernike Institute for
Advanced Materials, University of Groningen, The Netherlands}
\date{July, 3rd 2008}

\begin{abstract}
Metastable manganite perovskites displaying the antiferromagnetic so-called E-phase are predicted to be multiferroic. Due to the need of high-pressures for the synthesis of this phase, this prediction has only been confirmed in bulk HoMnO$_3$. Here we report on the growth and characterization of YbMnO$_3$ perovskite thin films grown under epitaxial strain. Highly-oriented thin films, with thickness down to ~30nm, can be obtained showing magneto-dielectric coupling and magnetic responses as those expected for the E-phase. We observe that the magnetic properties depart from the bulk behavior only in the case of ultrathin films (d $<$ 30nm), which display a glassy magnetic behavior. We show that strain effects alone cannot account for this difference and that the film morphology plays, instead, a crucial role.
\end{abstract}

\pacs{75.70.Ak, 75.80.+q, 75.50.Ee}
\maketitle

The search for materials combining two or more ferroic orderings -the so-called multiferroics- has boosted in recent years\cite{Hil00,Fie05,Eer06}. (Anti)ferromagnetism and standard ferroelectricity have been shown to be mutually exclusive\cite{Hil00}; thus ferroelectricity in multiferroics should be associated to non-standard mechanisms. A lot of attention has been drawn towards AMnO$_3$ (A=Tb, Dy) orthorhombic manganites\cite{Kim03,Kim03b}, where the ferroelectric displacements have their origin in the Dzyaloshinskii-Moriya interaction\cite{Ser06,Mal08,Xia08} and are directly induced by a spiral antiferromagnetic ordering\cite{Ken05,Mos06}. As magnetism and ferroelectricity are intimately related in these compounds, the magnetoelectric coupling is very strong and allows the magnetic control of the polarization\cite{Kim03}.

Less attention has been paid to orthorhombic manganites with smaller A cations (A=Lu,Yb,Tm,Er,Ho). The stable structure of these manganites is hexagonal, but the metastable orthorhombic perovskite phase can be stabilized by means of high pressure synthesis\cite{Woo73} or epitaxial growth\cite{Sal98}. The Mn-lattice of these compounds displays a magnetic transition to an inconmensurate antiferromagnetic structure at T$_N$$\sim$42K, followed by a lock-in transition to a commensurate E-type antiferromagnetic ordering at lower temperatures\cite{Tac07}. Recent theoretical work\cite{Ser06,Pic07} suggests that the E-type magnetic structure should induce ferroelectricity with large polarization values (0.5-12$\mu$C/cm$^2$). Up to date, only Lorenz et al.\cite{Lor04,Lor07} have shown, by means of dielectric and pyroelectric current measurements, the existence of a strong magnetodielectric coupling and a spontaneous polarization in the HoMnO$_3$ manganite. Dielectric and magnetoelectric characterization of other E-phase compounds, in order to confirm the theoretical predictions, is still lacking due to the difficult synthesis of this metastable phase. Moreover, more generally, despite their interest in applications, thin films of orthorhombic magnetoelectric manganites have been rarely reported.

Here we report on the growth, structural, magnetic and electrical characterization of YbMnO$_3$ perovskite thin films. YbMnO$_3$ (YbMO) thin films were deposited on (111)-SrTiO$_3$ (STO) substrates by Pulsed Laser Deposition (PLD) using a KrF excimer laser with $\lambda$=248nm. The deposition was performed at 750 $^o$C in an oxygen pressure of 0.15 mbar, with a laser fluence of $\sim$2 J/cm$^2$ and a laser repetition rate of 1 or 2 Hz. After deposition, the films were cooled down (-5$^o$C/min) to room temperature under an oxygen pressure of 100 mbar. The crystallinity and structure of the films was studied by both standard X-ray diffraction (XRD)($\lambda$=1.540$\AA$) and synchrotron radiation at the W1 beamline of HASYLAB ($\lambda$=1.393$\AA$). Cross-sections of the thin films on the substrates were analyzed using transmission electron microscopy (TEM) (JEOL 2010F). The thickness of the films -determined by modeling the XRD reflectivity- ranged between 7 and 64.5nm. The growth rate was estimated as ~0.065$\AA$/pulse. Magnetic characterization was performed by means of a Quantum Design SQUID (MPMS), while for the dielectric measurements a capacitance bridge (Agilent 4284A) hooked to a Quantum Design PPMS was used.

The bulk orthorhombic perovskite structure of the YbMnO$_3$ perovskite belongs to the Pbnm space group and has cell parameters a=5.2208$\AA$, b=5.8033$\AA$ and c=7.3053$\AA$\cite{Hua06}. The SrTiO$_3$(111) substrate (cubic perovskite with a=3.905$\AA$ ) can accommodate the YbMnO$_3$ structure with the two following orientations relationships: (i) YbMnO$_3$[101]//SrTiO$_3$[111] (out of plane) and YbMnO$_3$[101]//SrTiO$_3$[11$\overline{2}$] (in plane); and (ii) YbMnO$_3$[011]//SrTiO$_3$[111] (out of plane) and YbMnO$_3$[011]//SrTiO$_3$[11$\overline{2}$] (in plane). In the first case, the in-plane lattice mismatch (u= 1- (a$_{film}$/a$_{sub}$)) is tensile (6.4 $\%$) along STO[11$\overline{2}$] and compressive (-5$\%$) along STO[1$\overline{1}$0], while in the second case the strain is tensile in both directions (2.7$\%$ and 5.5 $\%$, respectively). The out-of-plane [101] orientation is more likely since the opposite sign of the lattice mismatch in both in-plane directions allows a better strain accommodation.

\begin{figure}
  \includegraphics[scale=0.35]{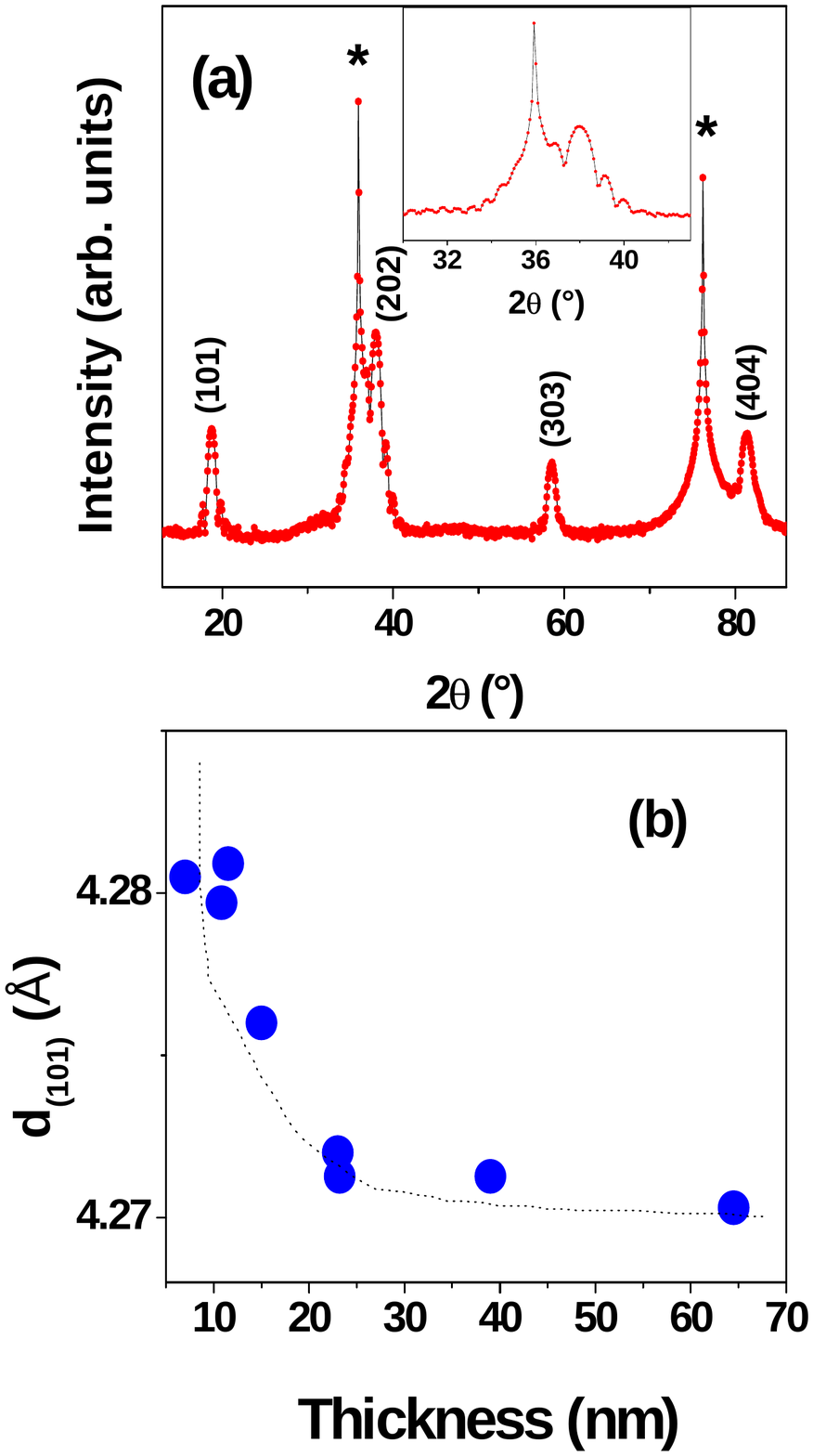}
  \caption{(Color online)(a) X-ray synchrotron $\theta-2\theta$ spectra corresponding to an 11.5nm YbMnO$_3$ film grown on (111)-SrTiO$_3$. The blow-up in the inset shows the presence of thickness oscillations; (b) Evolution of the out-of-plane interplanar distance (d$_{101}$) as a function of the thickness for different YbMO films.}
\end{figure}

Figure 1(a) shows a synchrotron XRD pattern recorded on a 12nm thick thin film. The film is shown to be single phase and, as expected, (101)-oriented. Figure 1(b) collects the evolution of the out-of-plane interplanar d$_{101}$ distance as a function of the thickness. In all cases the obtained values are larger than the bulk value (4.25$\AA$), indicating that the films grow under an overall compressive strain. The decrease of d$_{101}$ when increasing thickness indicates a progressive strain relaxation, probably through the formation of misfit dislocations. This is reflected in the increment of the diffuse component of the rocking curves recorded around the (202) reflections (not shown here), whose full-width-half-maximum increases from 0.09$^o$, for a 11.5 nm thick film, to 0.28$^o$, for a 39nm thick film.

\begin{figure}
  \includegraphics[scale=0.3]{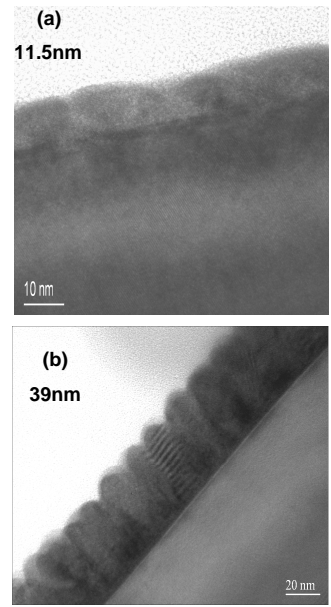}
  \caption{Cross-sectional transmission electron microscopy images corresponding to 11.5 nm (a) and 39nm (b) thick YbMnO$_3$ thin films.
}
\end{figure}

Figure 2(a) shows The TEM cross-section of a 11.5nm thick thin film, showing a relatively flat surface. This is confirmed by atomic force microscopy (AFM) images (not shown here), which display a typical root mean square roughness (RMS) lower that 0.4nm and an overall height variation of less than 1nm. Upon increasing thickness, the films become rougher. Figure 2(b) presents the TEM cross-section for a 39nm thick thin film, showing the presence of columnar growth with relatively deep cusps at the grain boundaries. The AFM images of the same film present a RMS roughness of 0.7nm and an overall height variation of 4nm. The increase of the roughness with the thickness is typical of a 3D growth mechanism.

In order to probe the Mn valence state of the films, we have measured the Mn-3s x-ray photoemission spectra (XPS) of films with different thickness. The Mn-3s level displays an energy splitting originated by the intra-atomic exchange coupling between 3s and 3d electrons; therefore, the magnitude of the splitting has been proposed to be proportional to the local Mn valence\cite{Gal02}. Figure 3 shows the spectra obtained for two TMO films with thickness of 11.5nm and 39nm. Both films display a splitting of $\sim$5.29eV, which is within the range reported for Mn ions with a 3+ nominal valence\cite{Gal02}. From these measurements it can be inferred that the stoichiometry of our films is consistent with the nominal one and, importantly, there are no composition stoichiometry gradients when varying the thickness.

\begin{figure}
  \includegraphics[scale=0.3]{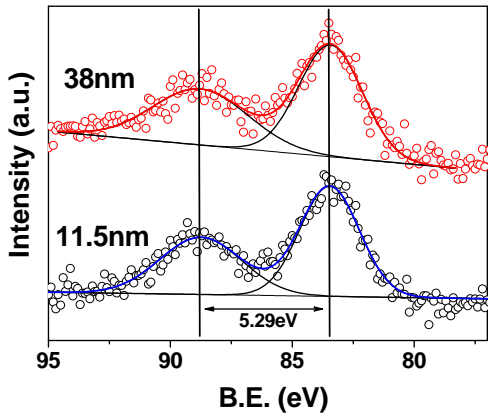}
  \caption{(Color online) Mn3s X-ray photoemission spectrum corresponding to 11.5 and 38nm YbMnO$_3$ thin films. The backgrounds are modeled by means of a Shirley function.}
\end{figure}

Figure 4(a) shows the evolution of the magnetization of a 35nm film as a function of temperature. It is found that the magnetization, largely dominated by the presence of the  strongly paramagnetic Yb ions, follows a Curie-Weiss law, as can be better appreciated in the linear evolution of the inverse susceptibility with temperature (inset of Figure 4(a)). The extracted effective magnetic moment is $\mu$$_{eff}$$\sim$(8$\pm$1)$\mu$$_B$, which is in reasonable agreement with the expected moment for a combination of Mn$^{3+}$ and Yb$^{3+}$ ions (6.7$\mu$$_B$). The extrapolated Curie-Weiss temperature is -25K, showing the presence of antiferromagnetic interactions. We notice that in bulk measurements, due to the paramagnetic contribution of Yb, the N\'{e}el temperature (43K) is only reflected as a very subtle feature in the magnetic susceptibility\cite{Hua06}. In thin films, where the magnetic signal is rather small and the signal-to-noise level is considerably lower than in bulk measurements, such a subtle feature can be easily missed and it is not surprising that it does not show in our data.

\begin{figure}
  \includegraphics[scale=0.3]{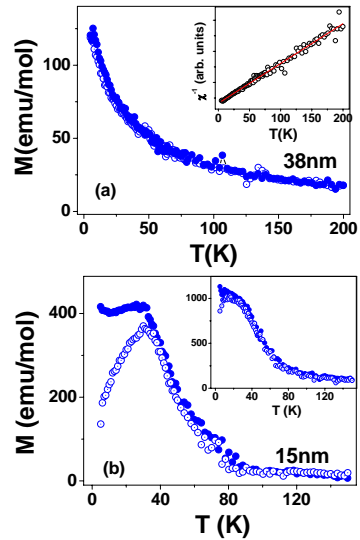}
  \caption{(Color online)(a) Zero field and field cooled (1kOe) magnetization as a function of the temperature corresponding to a 38nm YbMO thin film on a (111) STO substrate. The inset shows the evolution in temperature of the inverse susceptibility $\chi$$^{-1}$(T). The full line shows a fitting assuming a Curie-Weiss behavior; (b) Zero field and field cooled magnetization as a function of the temperature of a 15nm film under fields of 1kOe (main Panel) and 4kOe (Inset). In all cases, the field was applied in the plane of the film.}
\end{figure}

The magnetic behavior is remarkably different in ultra-thin films, as shown in Figure 4(b) for a 15nm film measured under a 1kOe field applied in-plane. Several features can be observed: field-cooled and zero-field-cooled measurements split below $\sim$40K, which is close to the magnetic ordering temperature of the bulk material. In addition, the zero-field-cool run shows a cusp at $\sim$30K. If the measurement is repeated under a higher field (4kOe, inset of Figure 4(b)), both the splitting temperature and the cusp shift to lower temperatures. From these features the presence of competing ferromagnetic/antiferromagnetic interactions leading to a glassy behavior can be suggested. As this behavior has  been observed only in ultrathin films ($<$ 30nm), it is tempting to attributed it to a strain effect; however, it should be noted that, in addition to be more strained (see Figure 1(b)), ultrathin films also display a much flatter morphology (see Figure 2). Interestingly, we find that if we modify the growth conditions to obtain rougher ultrathin films (RMS roughness$>$ 0.6nm), the magnetic behavior becomes Curie-Weiss like. This strongly indicates that the glassy behavior is related to a flatter morphology and shows that a more complex scenario than purely strain effects (involving surface or finite size effects) should be taken into account.

In order to measure the electrical properties, 15 and 35nm thick YbMnO$_3$ films were also grown on top of SrTiO$_3$(111) with a 20nm thick SrRuO$_3$ (SRO) conductive buffer layer. The optimization of the growth of the bottom electrode will be reported elsewhere\cite{Rub08}. The SRO buffer layer was found to be fully strained with the substrate, while the obtained out-of-plane YbMnO$_3$ d$_{101}$ distances were $\sim$4.28$\AA$ for 15 nm films and $\sim$4.27$\AA$ for 35 nm films. This is in agreement with those in Figure 1(b), indicating that the films grown on a SRO buffer are equally strained that those grown on bare STO(111). The roughness of the SRO/YbMnO films ranged between 0.5 and 0.9nm. Either SrRuO$_3$ or Au were used as top electrodes.

\begin{figure}
  \includegraphics[scale=0.3]{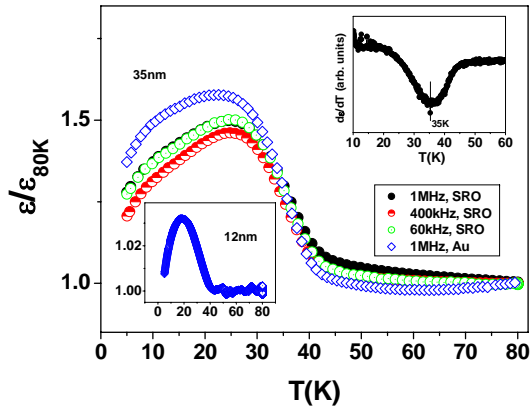}
  \caption{(Color online) Normalized dielectric constant $\epsilon$ as a function of temperature (T), for different frequencies, corresponding to two different YbMnO$_3$ films with thickness of 35nm. Either SrRuO$_3$ or Au were used as top electrodes. (Lower-left inset) Normalized $\epsilon$ vs. T for an ultrathin 12nm YbMnO$_3$ film. In the case of those films with Au top electrodes, a linear decrease of the $\epsilon$ was found above $\sim$40K and thus, in order to compare them with the SrRuO$_3$ electrode data, a linear background has been  subtracted from the raw data; (Upper-right inset) First derivative of $\epsilon$ with respect to T, as a function of T, of a 35nm YbMnO$_3$ film at a frequency of 1MHz.}
\end{figure}

The evolution of the dielectric constant $\epsilon$ (normalized by its 80K value) as a function of the temperature is shown in the main panel of Figure 5, at different frequencies, for two 35nm YbMnO$_3$ films with different top electrodes (SrRuO$_3$ and Au). A clear up-turn of $\epsilon$ at $\sim$42K is observed, which is in excellent agreement with the N\'{e}el temperature of the bulk compound. The temperature evolution of the first derivative of the dielectric permitivity (d$\epsilon$/dT), evidences the presence of a minimum at 35K. This temperature of largest slope of $\epsilon$(T) has been previously shown to correspond to the lock-in transition in HoMnO$_3$ and YMnO$_3$ bulk samples\cite{Lor04}. In our case the minimum of d$\epsilon$/dT is also in good agreement with the transition to the E-phase in bulk YbMnO$_3$(35K)\cite{Tac07}. The appearence of features in $\epsilon$(T) at the magnetic ordering temperatures indicates the presence of a magneto-dielectric effect, which has not been previously reported in YbMnO$_3$. Figure 5 also shows that $\epsilon$ reaches a maximum at $\sim$25K and further decreases for lower temperatures. This behavior resembles the case of HoMnO$_3$ and YMnO$_3$ (Refs.\cite{Lor04,Mar07}) and is not fully understood. It is important to notice that all the features described above were found to be sample, top-electrode and frequency independent in the range 50kHz-1MHz, indicating that they reflect the intrinsic electric behavior of the films\cite{Cat06}. This is so despite the large observed differences in the absolute values of the measured $\epsilon$, which strongly increases at low frequencies. This effect, which is not reflected in the normalized data, is due to an important contribution from grain boundaries and interfaces, and to the large dielectric losses observed.

The lower-left inset of Figure 5 shows the normalized $\epsilon$ as a function of the temperature for an ultrathin 12nm film. This film displays qualitatively the same features as the thicker 35nm films; however, the magneto-dielectric effect seems to be severely reduced. Thinner films displayed smaller -although still considerable- losses ($\tan\delta\sim$3 for 35nm films and $\tan\delta\sim$1 for 12nm films at 1MHz) and reduced electrical conductivities compared to their thicker counterparts ($\sim$0.5mS for 35nm films and $\sim$0.03mS for 12nm films). This is in agreement with a lesser contribution from grain boundaries, which have a large associated conductivity, in the ultrathin films and it is consistent with the observed differences in morphology. Any correlation between the reduced magneto-dielectric effect and the glassy magnetic behavior of ultrathin films would be highly speculative. Unfortunately, the presence of high losses also prevents to directly probe the presence of ferroelectricity by means of pyroelectric or ferroelectric loops measurements. Future efforts have to concentrate in the increase of the resistivity by controlling the film microstructure.

To conclude, we have grown pulsed laser deposited films of YbMnO$_3$ perovskite. Magnetodielectric coupling was observed for the first time in this metastable perovskite. Films thicker than 30nm showed magnetic properties consistent with the bulk behavior, while the magnetic properties depart from the bulk case for ultrathin films with atomically flat surfaces. Results reported here reinforce the importance of E-phase compounds in the search of magnetoelectric materials and show that epitaxial growth is a useful tool for the synthesis of these metastable phases.

We would like to thank U. Adem, G. Catalan and Nandang Mufti for useful discussions and T.F. Landaluce and P. Rudolf for the assistance with and access to the XPS. This work was supported by the E.U. STREP MaCoMuFi (Contract FP6-NMP3-CT-2006-033221)

\end{document}